# MIMO *In Vivo*

Chao He, Yang Liu, Thomas P. Ketterl, Gabriel E. Arrobo, and Richard D. Gitlin
Department of Electrical Engineering
University of South Florida
Tampa, Florida 33620, USA
Email: {chaohe, yangl}@mail.usf.edu, {ketterl, garrobo, richgitlin}@usf.edu

*Abstract*—We present the performance of MIMO for *in vivo* environments, using ANSYS HFSS and their complete human body model, to determine the maximum data rates that can be achieved using an IEEE 802.11n system. Due to the lossy nature of the *in vivo* medium, achieving high data rates with reliable performance will be a challenge, especially since the *in vivo* antenna performance is strongly affected by near field coupling to the lossy medium and the signals levels will be limited by specified specific absorption rate (SAR) levels. We analyzed the bit error rate (BER) of a MIMO system with one pair of antennas placed *in vivo* and the second pair placed inside and outside the body at various distances from the *in vivo* antennas. The results were compared to SISO simulations and showed that by using MIMO *in vivo*, significant performance gain can be achieved, and at least two times the data rate can be supported with SAR limited transmit power levels, making it possible to achieve target data rates in the 100 Mbps.

## I. Introduction

One appealing aspect of the emerging *Internet of Things*, is to consider *in vivo* networking as a rich application domain for wireless technology in facilitating wirelessly enabled healthcare. Wireless technology has the potential to synergistically advance healthcare delivery solutions by creating new science and technology for *in vivo* wirelessly networked cyber-physical systems of embedded devices.

*In vivo* wireless networks have certain characteristics such as low-complexity, limited transmission and processing power, reduced latency, high reliability, and operation in a highly lossy and dispersive radio frequency (RF) channel [1], and potential near-field operation [2]. It is the purpose of this paper to demonstrate that owing to the highly dispersive nature of the *in vivo* channel, achieving stringent performance requirements will be facilitated by the use of multiple-input multiple-output (MIMO) communications [3] to achieve enhanced data rates. For example, one application for MIMO *in vivo* communications is wirelessly transmitted low-delay High Definition video during Minimally Invasive Surgery (MIS) [4].

In this paper, we present advances in modeling MIMO *in vivo* wireless communications for embedded devices of limited complexity and power, meeting the high bit rate and low latency requirements of many surgical applications (e.g., HD video) and is organized as follows: In section II, we summarize the prior work on MIMO technology, the *in vivo* environment and MIMO for wireless body area networks (WBANs). In section III we describe our approach, to MIMO *in vivo* communications. Section IV presents the simulation setup and results of the performance of MIMO *in vivo*. Finally, in Section V we present our conclusions and future research directions.

## II. Literature Review

### A. MIMO technology

MIMO technology, the use of multiple antennas both in transmitter and receiver, can significantly improve the capacity and performance of the communication system in comparison to the conventional system with single antenna. In modern communications systems, the combination of MIMO and OFDM technology [5] is extremely popular and takes advantage of multipath and materially improves the radio transmission performance.

### B. In vivo wireless communications

Understanding and optimizing the wireless *in vivo* channel, which is critical to advancing many bio-medical and other procedures, is an exciting new communications environment. The research objectives of the USF team have been directed towards increasing the reliability and the communications efficiency for the *in vivo* channel. The authors in [1] performed signal strength and channel impulse response simulations using an accurate human body model and investigated the variation in signal loss at different RF frequencies as a function of position around the human body.

### C. MIMO in WBAN

There has been some research that focuses on MIMO WBANs. There are a few models for MIMO system that can be applied to WBANs. In [6], the authors place the antennas on human clothing and analyze the performance of the proposed wearable MIMO systems, which has a significantly better performance than the previous system on a handheld platform. The wideband body-to-body radio channel in MIMO system is investigated in [7] and the authors also present several critical characterizations of the channel such as path loss, body shadowing, and small-scale fading. To the best of our knowledge, MIMO system for *in vivo* environments have not been studied in the literature, and due to the tremendous opportunity to create novel applications (e.g. transmitting HD video from inside the human body) for the *in vivo* environment, in the following section, we present a study of the performance of MIMO system in the *in vivo* environment.

## III. MIMO *In Vivo*

A MIMO *in vivo* system may be required to achieve high data rates using low transmission power to comply with the SAR requirements.

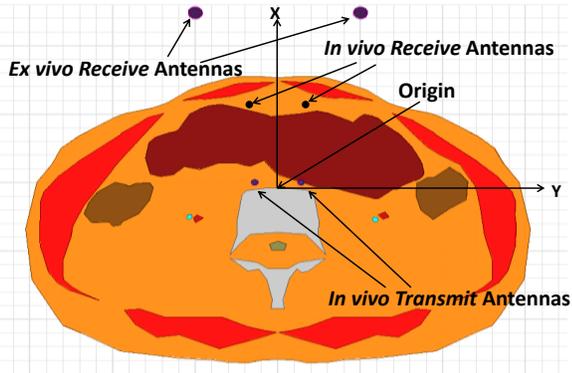

Fig. 1. Antenna simulation setup showing locations of the MIMO antennas. SISO antenna placement, not shown, is given in Table 1.

TABLE I. LOCATIONS OF THE *EX VIVO* (YELLOW) AND *IN VIVO* (GREEN) ANTENNAS WITH RESPECT TO THE ORIGIN (X=0, Y=0) SHOWN IN FIG. 1

| Simulation Scenario | MIMO | | | | SISO |
|---|---|---|---|---|---|
| | Antenna 1 | | Antenna 2 | | Antenna |
| | *X (mm)* | *Y (mm)* | *X (mm)* | *Y (mm)* | *X (mm)* |
| 1 | 130 | 50 | 130 | -50 | 130 |
| 2 | 100 | 50 | 100 | -50 | 100 |
| 3 | 70 | 30 | 70 | -30 | 70 |

The wireless *in vivo* channel is an exciting and challenging new environment that has not been well documented in the literature beyond limited analyses of signal attenuation and shadowing of human tissues limited to the Medical Implant Communication Service (MICS) frequency band (from 402 to 405 MHz). The IEEE P802.15 TG6 WBANs channel model provides guidance as to how the channel model should be developed for body area networks. For *in vivo* communications, however this document is limited and only provides the path loss exponent.

In the far-field electromagnetic waves behave as plane waves and the total radiated power does not change with changing radial distance from the antenna since the distance between transmitter and receiver is large and free space is considered lossless, as in cellular networks. However, since the *in vivo* antennas are radiating into a complex lossy medium, the radiating near field will strongly couple to the lossy environment [8]. This means the radiated power is strongly dependent on radial and angular position and that near field effects will always have to be taken into account when operating in the *in vivo* environment. In the radiating near field, the electric and magnetic fields behave differently compared to the far field. Hence, the wireless channel inside the body requires different link equations. Additionally, since the wavelength of the signal is much longer than the propagation environment in the near field, the delay spread concept and multi-path scattering of cellular network is not directly applicable to channels inside the body. This will directly affect the correlation between antennas, which is critically important for MIMO performance.

The achievable transmission rates in the *in vivo* environment have been simulated using a model based on the IEEE 802.11n standard since this OFDM-based standard supports up to 4 spatial streams (4x4 MIMO). Because of the form factor constraint inside the human body, our initial study is restricted to 2x2 MIMO. Also, the standard allows different modulation and coding schemes (MCS) that are represented by a MCS index value and use either 20 MHz or 40 MHz bandwidth. Due to the target data rates for the *MARVEL CM* (~80–100 Mbps), the MCS index values of our interest for MIS HD video applications are from 13 for 20MHz and 10 for 40MHz. The following section presents the simulation setup for the communications system and the *in vivo* environment.

## IV. SIMULATION SETUP AND RESULTS

The simulations for the electromagnetic wave propagation were performed in ANSYS HFSS 15.0.3 using the ANSYS human body model. The model consists of a detailed adult male with over 300 muscles, organs, and bones with a geometrical accuracy of 1 mm and realistic frequency dependent material parameters (conductivity and permittivity) from 20 Hz to 20 GHz. The antennas used in the simulations were monopoles designed to operate at the 2.4 GHz ISM band in their respective medium; free space for the *ex vivo* antennas and inside the body for *in vivo* antennas. We choose monopoles due to their smaller size, simplicity in design and omni-directionality. For the *in vivo* case, the monopole's performance and radiation pattern will vary with position and orientation inside the body [1], [8], making the performance of the *in vivo* antenna strongly dependent on the antenna type.

As shown in Fig. 1, two *in vivo* antennas are placed inside the abdomen to simulate placement of transceivers in laparoscopic and intestinal medical applications. The two *ex vivo* antennas are placed at varying locations around the body at the same planar height as the *in vivo* antennas. The locations with respect to the *in vivo* antennas are given in Table I. For the performance comparison with SISO *in vivo* systems, the locations of the *ex vivo* antennas for SISO cases are also listed in Table I. The two *in vivo* antennas are located 14 cm from either side of the origin along the Y axis, while in SISO cases, the single *in vivo* antenna is located at the origin.

To evaluate the BER performance of *in vivo*, we set up an OFDM-based (IEEE 802.11n) wireless transceiver model operating at 2.4 GHz with varying MCS index value, and bit rates in Agilent SystemVue for different *in vivo* channel setups in HFSS. The system block diagram is shown in Fig. 2. The transmission power is set to be 0.412mw and the thermal noise power is set to -101 dBm. The transmission power used in this investigation was derived by the authors in [9], which is the maximum power level that will assure the maximum allowable SAR levels are not exceeded.

Preliminary results have been obtained for a 2x2 MIMO setup with antennas operating at 2.4 GHz. From HFSS and the human body model, S-parameters between Tx and Rx antennas were extracted between 1 and 3 GHz. Then, the bit error rate (BER) and frame error rate (FER) for the IEEE 802.11n system were found by running 10,000 frames for each simulation for different MCS index values, for 20 MHz, for 800 ns guard interval, and different frame lengths. Fig. 3 shows the bit error rate (BER) as a function of the MCS index value for both SISO and MIMO *in vivo* cases. As observed in Fig. 3, the data rate at MCS index 10 can be supported by MIMO *in vivo* considering the worst case where *ex vivo* antennas are located at 13cm from *in vivo* antennas and the target BER below $10^{-6}$, while only data rate at MCS index 2 can be supported by SISO *in vivo* with the same distance between antennas as MIMO *in vivo* case. MCS equal to 10 and 2 correspond to 39 Mbps and 19.5 Mbps, respectively. We also discovered that as the *ex* and *in vivo* antennas distance becomes smaller, the performance gain becomes even bigger. At least two times data rate can be supported by MIMO *in vivo* compared with that of SISO *in vivo*. Fig. 4 shows the BER as a function of distance between the *ex* and *in vivo* antennas at different MCS index values equal to 11, 12, 13, 14 corresponding to 52, 78, 104, and 117 Mbps, respectively [10]. In the case when transmitting data at MCS equal to 13 (104 Mbps), the external antenna needs to be placed within 9.5 cm from the body to achieve a minimum BER of $10^{-3}$ and meet the requirement of at least 100 Mbps. So, for our application focus, it is possible to transmit high definition video with low latency from deep inside the human body during Minimally Invasive Surgery.

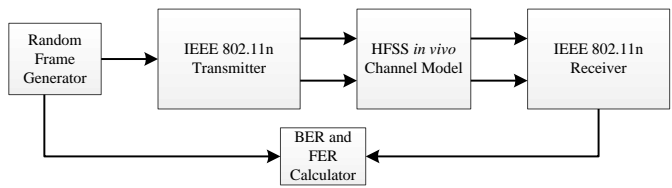

Fig. 2. Block diagram of system level simulation with HFSS *in vivo* channel model.

## V. CONCLUSION AND FUTURE RESEARCH

In this paper, we present an initial study of MIMO for *in vivo* environments to find out the maximum data rate that can be achieved in this challenging environment using a simulation method and results that utilizes accurate electromagnetic field simulations. We simulated a MIMO OFDM-based system that complies with the IEEE 802.11n standard. The simulations for the *in vivo* channel were obtained from HFSS that includes the human body model. From the preliminary data found in this study, as expected MIMO *in vivo* can achieve significant performance gains compared with SISO *in vivo*, and it is possible to achieve higher data rates (higher MCS index value) when the *in vivo* antennas are moved closer to the surface of the body.

Future research directions are to study the maximum transmission power that the *in vivo* antennas can use to comply with the SAR limits. Also *in vivo* channel will be characterized statistically and MIMO *in vivo* capacity will be studied theoretically, both incorporating near-field effects.

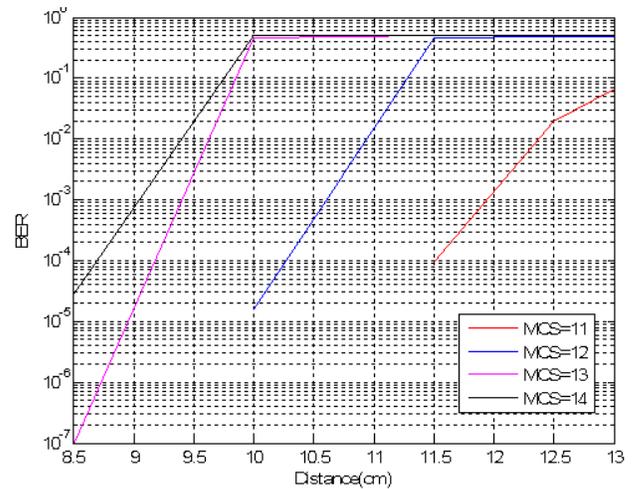

Fig. 4. MIMO *in vivo* BER performance comparison as function of the distance for different MCS Indexes.

## ACKNOWLEDGMENT

This research was supported in part by NSF Grant IIP-1217306, the Florida 21st Century Scholars program, and the Florida High Tech Corridor Matching Grants.

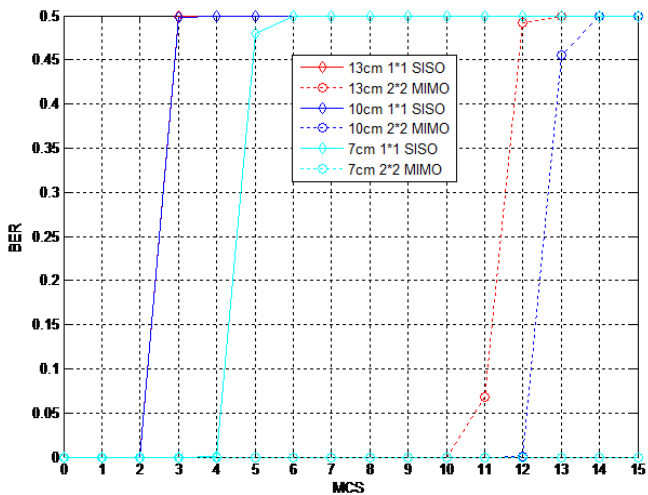

Fig. 3. SISO and MIMO *in vivo* BER performance comparison as function of the MCS index value.